 \documentclass[final,5p,times,twocolumn]{elsarticle}

\usepackage{amssymb}

\journal{Astroparticle Physics}

\begin{document}

\begin{frontmatter}



\title{Numerical simulations of diffusive shock acceleration in SNRs}


\author{V.N.Zirakashvili, V.S.Ptuskin}

\address{Pushkov Institute for Terrestrial Magnetism, Ionosphere and Radiowave
Propagation, 142190, Troitsk, Moscow Region, Russia}

\begin{abstract}
A new numerical model of the nonlinear diffusive shock acceleration is presented. It is
 used for modeling of particle acceleration in supernova remnants.
The model contains coupled spherically symmetric
hydrodynamic equations  and
the transport equations for energetic protons, ions and electrons. The forward and reverse shocks are
included in the consideration.
The spectra of
cosmic rays released into interstellar medium from a supernova remnant are determined.
The role of the reverse shock in the production of CR ions and positrons is discussed.
\end{abstract}


\end{frontmatter}



\section{Introduction}

The diffusive shock acceleration (DSA) process \cite{krymsky77,bell78,axford77,blandford78} is
considered as the principal mechanism for production of
galactic cosmic rays (CR) in supernova remnants (SNRs). A significant
theoretical progress in the investigation of this mechanism was
achieved
(see e.g. Malkov \& Drury
\cite{malkov01} for a review). However only during the last decade
the excellent results of X-ray and gamma-ray astronomy supplied
the observational evidence of the presence of multi-TeV energetic
particles in these objects (see e.g. Aharonian et al. \cite{aharonian08}).

Two shocks are produced  by super-sonically
moving supernova  ejecta after an explosion. A forward shock propagates
in the circumstellar medium while a reverse shock propagates
in the gas of ejecta. Generally it is believed that some part of thermal
particles is injected at the shock fronts into acceleration.

In this paper we present a new numerical model of nonlinear
diffusive shock acceleration. This model is a natural development
of the existing models \cite{berezhko94,kang06}. The solution of spherically symmetric
hydrodynamic equations is combined with the energetic particle
transport and acceleration at the forward and reverse shocks of a
supernova remnant. Nonlinear response of energetic particles via
their pressure gradient results in the self-regulation of
acceleration efficiency.

Our previous studies that used this model dealt with the CR spectra
produced by SNRs \cite{zirakashvili08b,ptuskin10}.
The input of the reverse shock was not taken into account there while
the acceleration by both shocks was considered for modeling of
non-thermal electromagnetic emission from the SNR RX J1713.7-3946 \cite{zirakashvili10}.

The paper is organized as follows.  The short description of the
model is given in Sect. 2. The results of modeling of evolution of
a supernova remnant in the interstellar medium are presented in
Sect. 3. Sect. 4 contains the discussion of our results. Our
conclusions are given in the last Section. The numerical code is described in Appendix.

\section{Nonlinear model of diffusive shock acceleration}

Hydrodynamical equations for the gas density  $\rho (r,t)$, gas velocity $u(r,t)$, gas pressure
$P_g(r,t)$, and the equation for isotropic part of the CR proton momentum distribution
 $N(r,t,p)$ in the spherically symmetrical  case are given by

\begin{equation}
\frac {\partial \rho }{\partial t}=-\frac {1}{r^2}\frac {\partial }{\partial r}r^2u\rho
\end{equation}

\begin{equation}
\frac {\partial u}{\partial t}=-u\frac {\partial u}{\partial r}-\frac {1}{\rho }
\left( \frac {\partial P_g}{\partial r}+\frac {\partial P_c}{\partial r}\right)
\end{equation}

\begin{equation}
\frac {\partial P_g}{\partial t}=-u\frac {\partial P_g}{\partial r}
-\frac {\gamma _gP_g}{r^2}\frac {\partial r^2u}{\partial r}
-(\gamma _g-1)(w-u)\frac {\partial P_c}{\partial r}
\end{equation}

\[
\frac {\partial N}{\partial t}=\frac {1}{r^2}\frac {\partial }{\partial r}r^2D(p,r,t)
\frac {\partial N}{\partial r}
-w\frac {\partial N}{\partial r}+\frac {\partial N}{\partial p}
\frac {p}{3r^2}\frac {\partial r^2w}{\partial r}
\]
\[
+\frac {\eta ^f\delta (p-p_{f})}{4\pi p^2_{f}m}\rho (R_f+0,t)(\dot{R}_f-u(R_f+0,t))\delta (r-R_f(t))
\]
\begin{equation}
+\frac {\eta ^b\delta (p-p_{b})}{4\pi p^2_{b}m}\rho (R_b-0,t)(u(R_b-0,t)-\dot{R}_b)\delta (r-R_b(t))
\end{equation}
Here $P_c=4\pi \int p^2dpvpN/3$ is the CR pressure, $w(r,t)$ is the  advective velocity of CRs,
$\gamma _g$ is the adiabatic index of the gas, and $D(r,t,p)$ is the CR diffusion coefficient.
It was assumed that the diffusive streaming of CRs results in the generation of magnetohydrodynamic (MHD)
waves. CR particles are scattered by these waves. That is why the CR advective velocity
 $w$ may differ from the gas velocity $u$. Damping of these waves results in an additional gas heating. It is
described by the last term in Eq. (3). Two last terms in Eq. (4)
correspond to the injection of thermal protons with momenta
$p=p_{f}$, $p=p_{b}$ and mass $m$ at the fronts of the forward and
reverse shocks at $r=R_f(t)$ and $r=R_b(t)$
respectively\footnote{We shall use indexes $f$ and $b$ for quantities
corresponding to the forward and
 reverse (backward) shock respectively.}. The
dimensionless parameters $\eta ^f$ and $\eta ^b$ determine the injection
efficiency.

The equation for ions is similar to Eq. (4). For ions with the
mass $M=Am$ and the mass number $A$ it  is convenient to use the
momentum per nucleon $p$ and the normalization of the ion spectra
$N_i$ to the baryonic number density. Then the number density of
ions $n_i$ is  $n_i=4\pi A^{-1}\int p^2dpN_i$. The ion pressure
$P_i=4\pi \int p^2dpvpN_i/3$ is also taken  into account in the
cosmic ray pressure $P_c$.

We shall neglect the pressure of energetic electrons. In other
words they are treated as the test particles. The evolution of
electron distribution is described by equation similar to Eq. (4)
with additional terms describing synchrotron and inverse Compton
(IC) losses.

CR diffusion is determined by magnetic inhomogeneities. Strong
streaming of accelerated particles changes medium properties in
the shock vicinity. CR streaming instability results in the high
level of MHD turbulence \cite{bell78} and even in the
amplification of magnetic field in young SNRs \cite{bell04}. Due
to this effect the maximum energy of accelerated particles may be
higher in comparison with previous estimates of Lagage and Cesarsky \cite{lagage83}.

According to the numerical modeling of this non-resonant instability,
the magnetic field is amplified by the flux of run-away highest energy
particles in the relatively broad region upstream of the shock \cite{zirakashvili08}.
Magnetic energy density is a small fraction
($\sim 10^{-3}$) of the energy density of accelerated particles.
This amplified almost isotropic magnetic field can be considered
as a large-scale magnetic field for lower energy particles which
are concentrated in the narrow region upstream of the shock.
Resonant streaming instability of these particles produces MHD waves
propagating in the direction opposite to the CR gradient. Strong nonlinear damping of
 these waves results in the gas heating (see the last term in Eq.(3)). CR 
gradient is negative upstream of the  forward shock and MHD waves propagate
in the positive direction. The situation changes downstream of the
forward shock where CR gradient is as a rule positive and MHD waves
propagate in the negative direction. This effect is mostly pronounced
downstream of the forward shock of SNR because the magnetic field is additionally
amplified by the shock compression and the Alfv\'en velocity
 $V_A=B/\sqrt{4\pi \rho }$ can be comparable with the gas velocity in the shock frame
 $u'=\dot{R}_f-u(R_f-0,t)$. As for CR diffusion coefficient, it is probably close to the Bohm value
$D_B=pvc/3qB$, where $q$ is the electric charge of
particles.

We apply a finite-difference method to solve Eqs (1-4) numerically
upstream and downstream of the forward and reverse shock (see Appendix).
A non-uniform numerical grid upstream of
the shocks at
 $r>R_f$ and $r<R_b$ allows to resolve small scales of hydrodynamical quantities appearing due to
the pressure gradient of low-energy CRs.
The gases compressed at the
forward and reverse shocks are separated by a contact
discontinuity at $r=R_c$ that is situated between the shocks.
An explicit conservative TVD scheme
\cite{trac03} for hydrodynamical equations (1-3) and uniform spatial grid were
used between the shocks.

Because of synchrotron losses
the spacial scale of the high-energy electrons can be rather small downstream of the shocks.  That is why
 we use a non-uniform numerical spacial grid for accelerated electrons downstream of the shocks.

The magnetic field plays no dynamical role in the model. Since we
have not perform the modeling of the amplification and transport
of magnetic field here, its coordinate dependence should be
specified for determination of cosmic ray diffusion and for
calculation of synchrotron emission and losses. We shall assume
below that the coordinate dependencies of the magnetic field and
the gas density coincide upstream and downstream of the forward
shock:
\begin{equation}
B(r,t)=B_0\frac {\rho }{\rho _0}\sqrt{\frac {\dot{R}^2_f}{M^2_AV_{A0}^2}+1}, \ r>R_c.
\end{equation}
Here $\rho _0$ is the gas density and $V_{A0}=B_0/\sqrt{4\pi \rho _0}$ is the Alfv\'en velocity
of the circumstellar medium. The parameter
  $M_A$ determines the value of the amplified magnetic field strength.
For low shock velocities $\dot{R}_f<M_AV_{A0}$ the magnetic field is not amplified.

The magnetic energy is about 3.5 percent
of the dynamical pressure
$\rho _0\dot{R}^2$ according to estimates from the width of X-ray
filaments in young SNRs \cite{voelk05}. This
number and characteristic compression ratio of a modified SNR shock
 $\sigma =6$ correspond to  $M_A\approx 23$. Since the plasma density $\rho $ decreases
towards the contact discontinuity downstream of the forward shock,
the same is true for the magnetic field strength according to Eq. (5). This
seems reasonable because of a possible magnetic dissipation in this region.

Situation is different downstream of the reverse shock at
$R_b<r<R_c$. The plasma flow is as a rule strongly influenced by
the Rayleigh-Taylor instability that occurs in the vicinity of the
contact discontinuity and results in the generation of MHD
turbulence in this region. We shall assume that the magnetic field
does not depend on radius downstream of the reverse shock while
the dependence in the upstream region is described by the equation
similar to Eq. (5):

\begin{equation}
B(r,t)=\sqrt{4\pi \rho _m}\frac {|\dot{R}_b-u(r_m)|}{M_A}
\left\{ \begin{array}{lll}
1, \ r<r_m,
\\
\rho /\rho _m, \\ r_m<r<R_b,
\\
\rho (R_b+0)/\rho _m, \\ R_b<r<R_c
\end{array} \right.
\end{equation}
Here $r_m<R_b$ is the radius where the ejecta density has a
minimum and equals $\rho _m$. This radius $r_m$ is generally close
to the reverse shock radius $R_b$ and is equal to it if the
reverse shock is not modified by the cosmic ray pressure.

\begin{figure}[t]
\includegraphics[width=8.0cm]{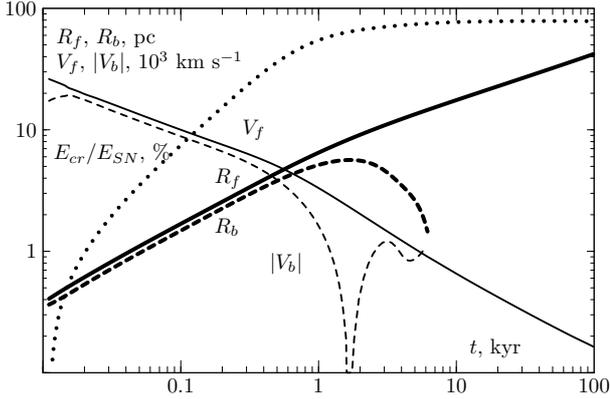}
\caption{Dependencies on time of the forward shock radius $R_f$ (thick solid line),
the reverse shock radius $R_b$ (thick dashed line), the forward shock velocity $V_f$
 (thin solid line) and the reverse shock velocity $V_b$
 (thin dashed line).
The ratio of CR energy and energy of supernova
explosion  $E_{cr}/E_{SN}$ (dotted line) is also shown.}
\end{figure}

\begin{figure}[t]
\includegraphics[width=8.0cm]{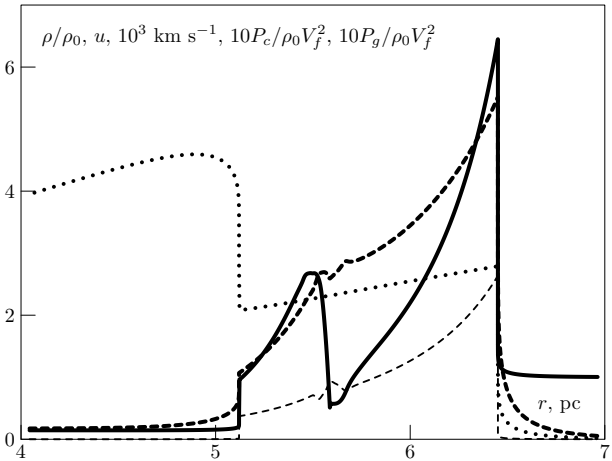}
\caption{Radial dependencies of the gas density (thick solid line), the gas
velocity (dotted line), CR pressure (thick dashed line) and the gas pressure (dashed line) at
 $t=10^3$ yr. At this epoch   the forward shock velocity is 3300 km s$^{-1}$, its radius is 6.5 pc,
the reverse shock velocity is 1650 km s$^{-1}$, its radius is 5.1 pc,
the magnetic
field strength downstream of the forward shock is 160 $\mu $G while the magnetic field strength
 downstream the reverse
shock is 56 $\mu $G. }
\end{figure}

CR advective velocity may differ from the gas velocity on the value of the radial
component of the Alfv\'en velocity
$V_{Ar}$ calculated in the isotropic random magnetic field:  $w=u+\xi _AV_{A}/\sqrt{3}$.
Here the factor $\xi _A$
describes the possible deviation of the cosmic ray drift velocity from the gas velocity.
The similar expression for the cosmic ray drift velocity is used upstream of
the reverse shock at $r<R_b$.
We shall use values $\xi _A=1$ and $\xi _A=-1$ upstream of the
forward and reverse shocks respectively, where Alfv\'en waves are
generated by the cosmic ray streaming instability and propagate in
the corresponding directions. The damping of these waves heats the gas
upstream of the shocks \cite{mckenzie82}
and limits the total compression ratios by a number close
to 6. This Alfv\'en heating produce a more efficient limitation of the shock modification
in comparison with the dynamical effects of the magnetic field considered by Caprioli et al. \cite{caprioli08} which
 are neglected in our study. 
In the downstream region of
the forward and reverse shock at $R_b<r<R_f$ we set $\xi _A=0$ and therefore $w=u$ in the major part
of this paper except the consideration of the Alfv\'en drift effects downstream of the shocks at the end
of Sect.3 and in Fig.10.

We shall use  the following diffusion coefficient

\begin{equation}
D=\eta _BD_B
\left\{ \begin{array}{lll}
\left( 1+\frac {M^2_AV^2_{A0}}{\dot{R}^2_f}\right) ^{g}\exp \left( \frac {r-R_f}{\xi _0R_f}\right) ,\ r>R_f,
\\
\left( 1+\frac {M^2_AV^2_{A0}}{\dot{R}^2_f}\right) ^{g},\ R_c<r<R_f,
\\
1, \ R_b<r<R_c, \\
\exp \left( \frac {R_b-r}{\xi _0R_b}\right) , r<R_b.
\end{array} \right.
\end{equation}

Here the parameter $g>0$ depends on  the type of
nonlinear wave damping which is relevant only for low velocity shocks 
$\dot{R}_f<M_AV_{A0}$ when the magnetic field is not amplified.
The parameter $\eta _B$ describes the possible deviations of
diffusion coefficient
from the Bohm value $D_B=vpc/3qB(r,t)$ for high-velocity shocks.
Since the highest energy particles are scattered by
small-scale magnetic fields,
their diffusion is faster than the Bohm diffusion \cite{zirakashvili08}.
The same is true for
smaller energy particles because they can be resonantly scattered only by a fraction of the magnetic
spectrum. We shall use the value $\eta _B=2$ throughout the paper.

We shall use the value of parameter  $g=1.5$ below. It corresponds
to the nonlinear wave damping  of the weak turbulence theory. Note
that a stronger Kolmogorov-type nonlinear damping used by Ptuskin
\& Zirakashvili \cite{ptuskin05} for estimate of maximum energy in
SNRs corresponds to $g=3$.

In real situation the level of MHD turbulence drops with distance
upstream of the shock and diffusion may be even faster there. This is 
qualitatively taken into account by the exponential factors in Eq.
(7). The characteristic diffusive scale of highest energy
particles is
 a small fraction $\xi _0<<1$ of the shock radius \cite{zirakashvili08}
and is determined by the generation and
transport of MHD turbulence in the upstream region \cite{vladimirov06,amato06}.
The value $\xi _0\sim \ln ^{-1}(D_c/D_s)$ is determined by ratio of diffusion
coefficient $D_c$ in the circumstellar medium and diffusion coefficient $D_s<<D_c$ in the vicinity of the
shock. The MHD turbulence is amplified exponentially in time before the shock arrival from the background
level by cosmic ray streaming instability.
 The characteristic range of $\xi _0$ is
$0.05\div 0.1$ \cite{zirakashvili08}.
We shall use the value $\xi _0=0.05$ below.

\begin{figure}[t]
\includegraphics[width=8.0cm]{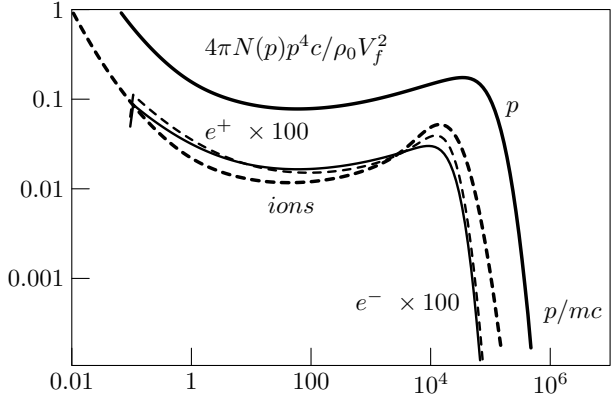}
\caption{Spectra of accelerated particles at $t=10^3$ yr. The
proton spectrum at the forward shock (thick solid), ion  spectrum
at the reverse shock (thick dashed), electron spectrum at the
forward shock (multiplied to the factor of 100, thin solid) and positron
spectrum at the reverse shock (multiplied to the factor of 100, thin dashed) are
shown. Spectrum of ions is
 shown as the function of momentum per nucleon and  normalized to the baryonic number density.}
\end{figure}

\begin{figure}[t]
\includegraphics[width=8.0cm]{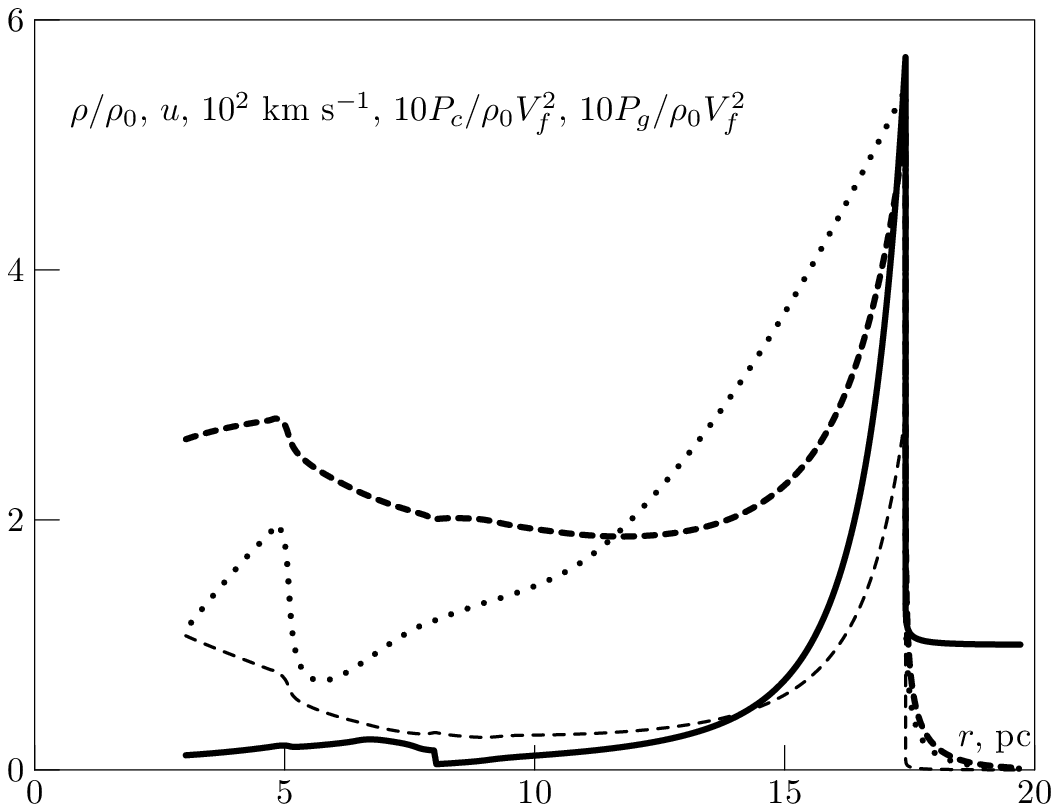}
\caption{Radial dependencies of the gas density (thick solid line), the gas
velocity (dotted line), CR pressure (thick dashed line) and the gas pressure (dashed line) at
 $t=10^4$ yr. At this epoch   the forward shock velocity is 660 km s$^{-1}$, its radius is 17 pc,
magnetic field strength downstream of the forward shock is 40 $\mu $G. }
\end{figure}

\begin{figure}[t]
\includegraphics[width=8.0cm]{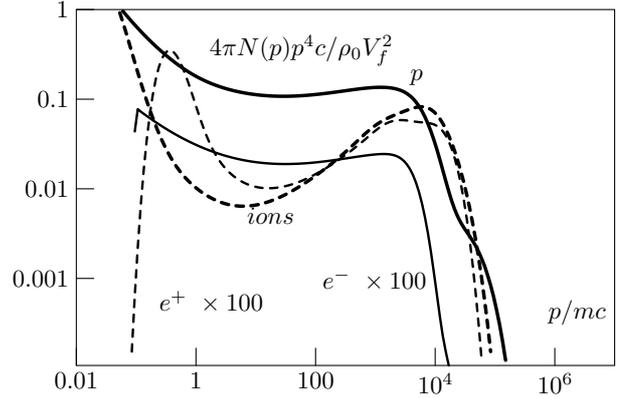}
\caption{Spectra of accelerated particles at $t=10^4$ yr.  The
spectrum of protons   (thick solid line) and electrons (multiplied to the factor of
 100, thin solid line) at the forward shock, ion spectrum (thick
dashed line) and positron spectrum (multiplied to the factor of 100,thin dashed
line) in the central part of the remnant are shown. Spectrum of
ions is shown as the function of momentum per nucleon and
normalized to baryonic number density.}
\end{figure}

It is believed that the supernova ejecta has some velocity distribution $P(V)$ just after
the supernova explosion \cite{chevalier82b}

\begin{equation}
P(V)=\frac {3(k-3)}{4\pi k}
\left\{ \begin{array}{ll}
1, \ V<V_{ej} \\
\left( V/V_{ej}\right) ^{-k}, \ V>V_{ej}.
\end{array} \right.
\end{equation}
Here the index $k$ characterizes the steep power-low part of this distribution.
The radial distribution of
ejecta density is described by the same expression with $V=r/t$. The characteristic ejecta
velocity $V_{ej}$ can be expressed in terms of energy of supernova explosion $E_{SN}$ and ejecta mass $M_{ej}$
as

\begin{equation}
V_{ej}=\left( \frac {10(k-5)E_{SN}}{3(k-3)M_{ej}}\right) ^{1/2}.
\end{equation}

\section{Numerical results}

Figures (1)-(9) illustrate the numerical results that are
obtained for the SNR shock propagating in the medium with a hydrogen number density
 $n_H=0.1$ cm$^{-3}$, magnetic field strength $B_0=5$ $\mu $G and temperature $T=10^4$ K.
The fraction $x_{He}=n_{He}/n_{H}=0.1$ of helium nuclei was assumed. The gas of ejecta does
not contain hydrogen in the case considered. 
We use the ejecta mass
 $M_{ej}=1.4M_{\odot }$,  the energy of explosion $E_{SN}=1.0\cdot 10^{51}$ erg and the parameter of ejecta
velocity distribution $k=7$. The value of the parameter $M_A=23$ was assumed.

The initial forward shock velocity is
 $V_0=2.9\cdot 10^4$ km s$^{-1}$. The injection efficiency is taken to be
independent on time $\eta ^b=\eta ^f=0.01$, and the injection
momenta are $p_{f}=2m(\dot{R}_f-u(R+0,t))$, $p_{b}=2m(u(R_b-0,t)-\dot{R}_b)$. Protons with a mass $m$ are
injected at the forward shock while ions with mass number $A$ and charge number $Z=A/2$
are injected at the reverse shock.
This high injection efficiency
results in the significant shock modification already at early stages of SNR expansion while the
thermal sub-shock compression ratio is close to 2.5 during the simulation. This is in agreement
with the radio-observations of young extragalactic SNRs
\cite{chevalier06} and with the modeling of collisionless shocks
\cite{zirakashvili07}. Similar values of the injection efficiency were found in hybrid modeling (see e.g.
  \cite{giacalone97}) and at the Earth bow shock in the solar wind \cite{ellison90}. 

As for the electron injection we assume a rather high injection
energy of electrons $E_{inj}=100$ MeV. This qualitatively
corresponds to some  models of suprathermal electron injection.
Partially ionized ions accelerated at the shocks up to relativistic
energies can produce multi-MeV electrons in the upstream region in
the course of photo-ionization by Galactic optical and infrared
radiation \cite{morlino11}. MeV electrons and
positrons are also present in the radioactive supernova ejecta
while gamma-rays from $^{56}$Co  decay in ejecta produce energetic
electrons via Compton scattering in the circumstellar medium \cite{zirakashvili10b}.
These
energetic particles may be additionally pre-accelerated via
stochastic acceleration in the turbulent upstream regions of the
shocks.

Below we assume that electrons are injected at the forward shock
with efficiency $\eta _-^f=10^{-3}\dot {R_f}^2/c^2$ while
positrons are injected at the reverse shock with efficiency $\eta
_+^b=10^{-6}$. These numbers are expected for the injection
mechanisms mentioned above (see Sect.4 for details). Since
electrons are considered as test particles our results can be
easily rescaled for any other injection efficiency. The chosen
injection rate at forward shock  maintains the electron to proton
ratio $K_{-p}$ of the order of $K_{-p}\sim 10^{-3}$ throughout the
simulation while time-independent
 positron injection at the reverse shock results in positron to ion ratio $K_{+i}$ increase from
$K_{+i}\sim 10^{-4}$ in
 the very beginning of SNR evolution up to $K_{+i}\sim 10^{-2}$ at several thousand years shortly before the
disappearance of the reverse shock.

In the real SNR the ions are also injected at the forward shock. We do not consider this process here
 in order to find the spectra of ions produced by the reverse shock. Because of the same reason
 we assumed the absence of hydrogen in the ejecta. Although this is the case for Ia/b/c and IIb supernovae
 it is not true for IIP supernovae. 
 It is expected that the
 spectra of ions injected at the forward and reverse  shock are similar to the spectra of protons.
 The production of
 CR ions at the forward shock was recently considered by Caprioli  et al. \cite{caprioli11}.

The dependencies on time of the shock radii $R_f$ and $R_b$, the
forward and reverse shock velocities $V_f=\dot{R}_f$ and
$V_b=\dot{R}_b$, CR energy $E_{cr}/E_{SN}$   are shown in Fig.1.
The calculations were performed until $t=10^5$ yr, when the value
of the forward shock velocity drops down to $\dot{R}_f=164$ km
s$^{-1}$ and the forward shock radius is $R_f=42$ pc.

At early times of SNR evolution the distance between reverse and
forward shocks is only 10$\% $ of the remnant radius. This is less
than 23$\% $ thickness for automodel Chevalier-Nadezhin solution
with $k=7$ \cite{chevalier82} and should be attributed to a strong
modification of both shocks by CR pressure. The reverse shock is
strongly decelerated only when the forward shock sweeps the gas
mass comparable with the ejecta mass at $t>100$ yr and when the
transition to the Sedov phase begins.

Radial dependencies of physical quantities at
 $t=10^3$ yr are shown in Fig.2. The contact discontinuity between the ejecta and the interstellar gas
is at $r=R_c=5.6$ pc. The reverse shock in the ejecta is situated at $r=R_b=5.1$ pc. At the Sedov stage
the reverse
shock moves in the negative direction and reach the center after seven thousand  years after the supernova
explosion. We stop the calculations in the region $r< R_b$ when the reverse shock radius $R_b<0.1R_f$.

\begin{figure}[t]
\includegraphics[width=8.0cm]{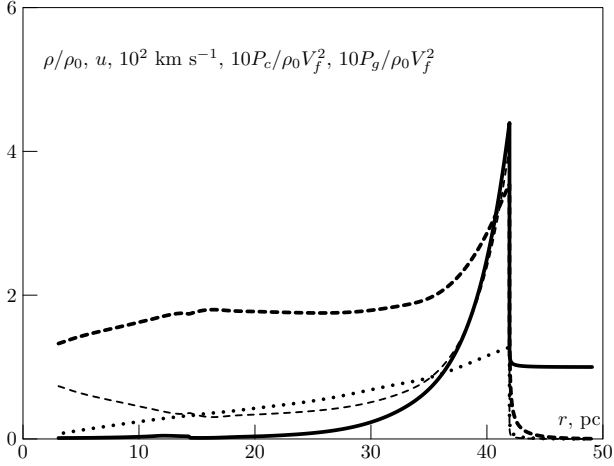}
\caption{Radial dependencies of the gas density (thick solid line), the gas
velocity (dotted line), CR pressure (thick dashed line) and the gas pressure (dashed line) at
 $t=10^5$ yr. At this epoch  the forward shock velocity is 164 km s$^{-1}$, its radius is 42 pc,
magnetic field strength downstream of the forward shock is 23 $\mu $G. }
\end{figure}

\begin{figure}[t]
\includegraphics[width=8.0cm]{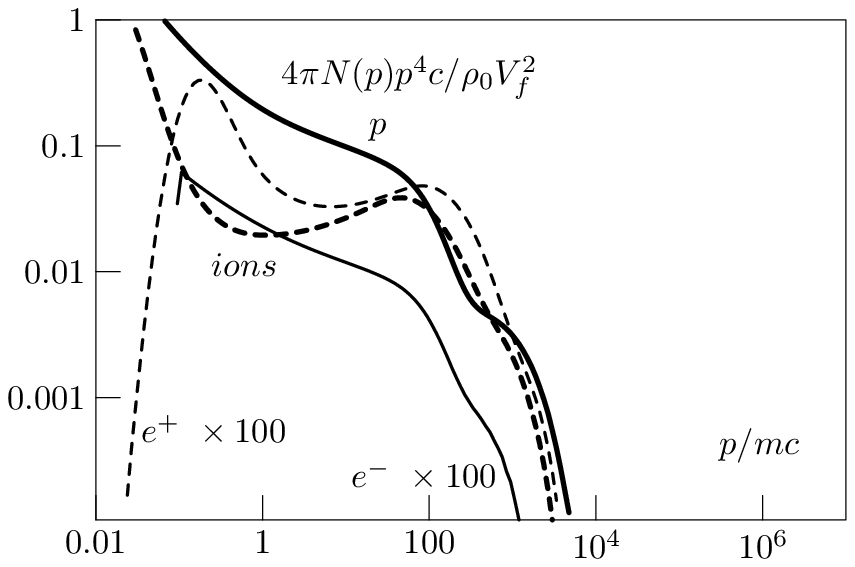}
\caption{Spectra of accelerated particles at $t=10^5$ yr.  The
spectrum of protons   (thick solid line) and electrons (multiplied to the factor of
 100, thin solid line) at the forward shock, ion spectrum (thick
dashed line) and positron spectrum (multiplied to the factor of
100, thin dashed line) in the central part of the remnant are
shown. Spectrum of ions is shown as the function of momentum per
nucleon and normalized to baryonic number density.}
\end{figure}

\begin{figure}[t]
\includegraphics[width=8.0cm]{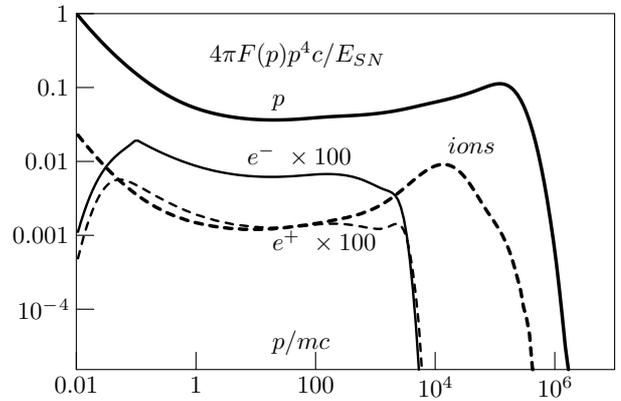}
\caption{ Spectra of particles produced in the supernova remnant
during $10^5$ yr. Spectrum of protons injected at the forward shock
(thick solid line ), spectrum of electrons injected at the forward
shock (thin solid line), spectrum of ions injected at the reverse
shock (thick dashed line) and the spectrum of positrons injected
at the reverse shock (thin dashed line) are shown. Spectrum of
ions is
 shown as the function of momentum per nucleon and  normalized to the baryonic number density.}
\end{figure}

\begin{figure}[t]
\includegraphics[width=8.0cm]{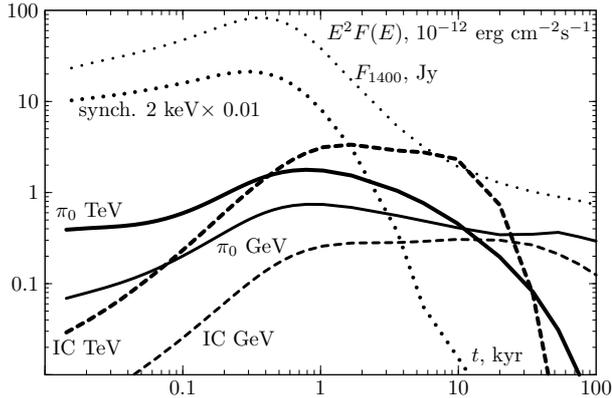}
\caption{Dependencies on time of gamma-ray fluxes from SNR at distance 4 kpc. We show
 the gamma-ray flux at 1 TeV produced via pion decay (thick solid line) and via IC process
 (thick dashed line), gamma-ray flux at 1 GeV produced via pion decay
 (thin solid line) and via IC process (thin dashed line).
 For the sake of simplicity only IC scattering of microwave background photons was considered. 
 The evolution of synchrotron X-ray flux at 2 keV (thick dotted line) and
 radio-flux at 1400 MHz $F_{1400}$  (dotted line) are  also shown.}
\end{figure}

\begin{figure}[t]
\includegraphics[width=8.0cm]{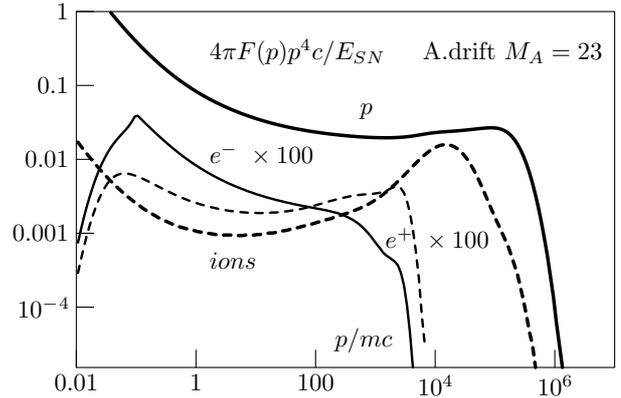}
\caption{ Spectra of particles produced in the supernova remnant
during $10^5$ yr in the model including the Alfv\'en drift
downstream of the shocks. Spectrum of protons injected at forward
shock (thick solid line ), spectrum of electrons injected at the
forward shock (thin solid line), spectrum of ions injected at the
reverse shock (thick dashed line) and the spectrum of positrons
injected at the reverse shock (thin dashed line) are shown.
Spectrum of ions is
 shown as the function of momentum per nucleon and  normalized to the baryonic number density.}
\end{figure}

It should be noted that our one-dimensional calculations cannot
adequately describe the development of the Rayleigh-Taylor
instability of the contact discontinuity. In the real situation
the supernova ejecta and the circumstellar gas are mixed by
turbulent motions in this region (see e.g. MHD modeling of Jun \&
Norman \cite{jun96}).

Spectra of accelerated protons and electrons at $t=10^3$ yr are shown in Fig.3.
At this epoch maximum energy of protons accelerated
in this SNR is about 100 TeV, while higher energy particles have
already left the remnant.

Radial dependencies of physical quantities at later epoch
 $t=10^4$ yr are shown in Fig.4. The reverse shock has reached the center of the remnant earlier. A
 weak reflected shock is clearly visible at $r=5$ pc.

The spectra of particles ar $t=10^4$ yr are shown in Fig.5.

Radial dependencies of physical quantities at the end of simulation at
 $t=10^5$ yr are shown in Fig.6. At this epoch the remnant is deeply in Sedov stage.
The contact discontinuity is at $r=14$ pc.

The spectra of particles  at $t=10^5$ yr are shown in Fig.7. The shock
 modification is not strong because the Alfv\'enic Mach number is close to 6 and the corresponding
 Alfv\'en heating upstream of the forward shock results in the lower compression ratio and acceleration
 efficiency. That is why the spectra of particles are steeper in comparison with ones at earlier epochs.

The spectra of particles produced during the whole evolution of the remnant are shown in Fig.8.
They are obtained
as the sum of the spectra integrated throughout simulation domain and  of the
time-integrated diffusive flux at the simulation
boundary at $r=2R_f$. At $t=10^5$ yr the maximum energy of currently accelerated particles drops down to
100 GeV because of nonlinear damping. Higher energy particles have already left the remnant.
Note that stronger
Kolmogorov-type damping with $g=3$ will result even in lower energies of the order of 1 GeV.
However we found that
the spectra do not change in this case.

We found that the maximum energy of CR protons is somewhat less than $10^{15}$ eV. It is
almost an order of magnitude
 lower in
 comparison with the results of Ptuskin et al. \cite{ptuskin10} where more optimistic assumptions
 $\eta _B=1$ and the spatially-uniform CR diffusion coefficient upstream of
the forward shock were used (cf. Eq.(7)).

Note that the synchrotron losses of run-away electrons and positrons
  were taken into account in our modeling.
The  cut-off energy of the leptonic spectra $\sim 5$ TeV shown in Fig.8 is determined
 by the magnetic field strength $B_0=5
\mu $G  in the  circumstellar medium and by the
 remnant age $t=10^5$ yr.

Evolution of non-thermal emission produced in the SNR at distance 4 kpc is shown in Fig.9.
It is worth noting that a significant part of the IC emission is produced
in the central region of the remnant
 at late epochs when the reverse shock have disappeared. This is because the magnetic field is
 rather weak in these regions.

\section{Discussion}

Although only about $5\% $ of supernova energy is transferred to
the particles accelerated at the reverse shock, they cannot be
neglected. First of all
 the ejecta has absolutely different composition  in comparison with
interstellar medium where the forward shock propagates. Since the
solar abundance corresponds to $1\% $ in the mass of heavy
elements while the ejecta
 can contain up to 50$\% $ of heavy elements it is clear that the reverse
shock will dominate in the production of heavy high-energy CR nuclei.

The relative input of the reverse shock at high energies is determined by
the relative energetics of the reverse and forward shocks that is of the order of 1/10 for
SNRs expanding in the uniform circumstellar medium. 

According to our results, approximately  $70\% $ of supernova
energy is transferred to  particles accelerated by forward shock.
It is significantly higher than the estimate of 10-20 $\% $ needed
to maintain CR density in the Galaxy if the supernova rate is 1/30
yr. One of possibilities to resolve this contradiction is the
assumption that CRs are accelerated only at small part of the
forward shock surface. This
 can be due to the dependence of the proton and ion injection on the shock
obliqueness \cite{berezhko04a}. This effect is observed in
SN 1006 and in the interplanetary medium.

This effect does not influence strongly the ion injection at the reverse shock.
It is expected that the random magnetic field is the main component of the field in the expanded ejecta.
This is because the magnetic field of ejecta originates from the magnetic field of the exploded star.
Thus the random magnetic field strength of the red super giant progenitor of IIP
supernovae is of the order
 of $10^4$ G similar to the magnetic field strength in the Sun interior
while the regular field is of the order of 1 G. After a homogenous
expansion from the initial stellar radius $10^{13}$ cm up to the radius $10^{19}$ cm
 of a young SNR  the frozen-in magnetic field drops
down to
 $10^{-8}$ G. Although this value is significantly lower than the
magnetic field in the interstellar medium
 it is strong enough for acceleration of particles up to 100 GeV in SNRs. A more realistic non-homogenous
expansion will result in the stretching of the field in the radial
direction. This can increase the magnetic field strength and ion
injection efficiency at the reverse shock. Magnetic fields can be
also amplified by non-resonant CR streaming instability suggested
by Bell \cite{bell04}.
If so the relative contribution of the reverse shock to the over-all CR spectrum  increases.

The effect also depends  on the type of the supernova explosion. It is known that
acceleration at the reverse shock occurs in Cas A SNR \cite{helder08,uchiyama08}. The
progenitor of the core-collapse Cas A supernova  had the radius
of the order of $10^{12}-10^{13}$ cm. On the other hand
white dwarfs that are  progenitors of
 Ia supernova explosions have small radii of the order of $10^9$ cm. Interior magnetic field $10^8$ G of
 the white dwarf will drop down to $10^{-12}$ G after the homogenous expansion of the young SNR.
 Such a weak magnetic field can result in the ineffective DSA at the reverse shock of Ia supernovae. This
 effect is probably observed in Tycho SNR (see Warren et al. \cite{warren05} for details).

Another possibility to suppress the CR production is related with the Alfv\'en drift downstream
of the forward shock \cite{zirakashvili08b}. It results in the steeper spectrum of CRs
accelerated at the forward shock. On the other hand the Alfv\'en
drift downstream of the reverse shock may produce even an opposite
effect because the CR gradient is positive in this region (see
Fig.2). As a result the input of the reverse shock will be
significant at TeV energies.

The over-all spectrum according to this model is shown in Fig.10. The value of $\xi _A=-1$ instead of
$\xi _A=0$ is used downstream of the shocks.
Because of the Alfv\'en drift the
positron spectrum at reverse shock is significantly harder than the electron spectrum at the forward shock.

It should be noted that the spectra of ions accelerated at the
reverse shock are harder than the proton spectra at the forward
shock (see Figs 3,8) in spite of the same level of the shock
modification for both shocks. This is because the shocks propagate
in the media with different properties. When the reverse shock
reach the flat part of the ejecta density distribution it
propagates in the medium with decreasing in time density. That is
why the number of freshly injected ions is low in comparison with
higher energy particles
 accelerated earlier. It results in the spectral hardening. This effect is absent at the forward shock
 propagating
 in the medium with a constant density. The Alfv\'en drift strengthens this effect (see Fig.10).

We adjust the electron (positron) injection to produce a sufficient
 number of electrons and positrons. The spectra shown in Figs 8 and 10
can explain CR Galactic  electrons and positrons. The expected positron injection
efficiency from the radioactive
decay of $^{44}$Ti is estimated as $\eta _+^b\sim M_{Ti}/(44M_{ej})\sim 10^{-6}$ \cite{zirakashvili10b}
for $^{44}$Ti mass
 of the order of $\sim 10^{-4}M_{\odot }$ as observed in SNRs \cite{renaud06}.
We used this number in our simulation.
As for the electron injection at the forward shock the relative
number of energetic electrons from photo-ionization
 of accelerated single charged He ions is of the order of
$\eta _-^f\sim x_{He}\gamma ^{-1}\ln ^{-1}(p_{max}/mc)\dot
{R_f}^2/c^2\sim 10^{-3}\dot{R}^2_f/c^2$. Here $\gamma \sim
I_{He}/\epsilon _{ph}\sim 10$ is the gamma-factor of
single-charged  He ion photo-ionized by galactic ultra-violet
photons with energy $\epsilon _{ph}\sim 10$ eV and $I_{He}=52$ eV
is the ionization potential of helium. This slightly overestimates
the electron injection
 in young SNRs where the acceleration is fast enough, $\gamma $ is closer to
 $\gamma \sim 100$ and the ionization is provided by eV optical photons. However we
 used this crude estimate that is justified in old SNRs for electron
injection in our simulations. This injection mechanism suggested
by Morlino \cite{morlino11} produces one order of magnitude higher
number of energetic electrons in comparison with the number $\eta
_-^f\sim 10^{-7}R^{-2}_{f,\rm{pc}}$ of Compton scattered electrons
energized by gamma-photons from
 $^{56}$Co radioactive decay in supernova ejecta \cite{zirakashvili10b}.
Here $R_{f,\rm{pc}}$ is the forward shock radius expressed  in parsecs.

The efficiency of electron (positron) acceleration at the reverse shock in SNRs can be observationally checked in
 the gamma-ray band. Young and middle-age ($t\sim 10^4$ yr) SNRs can be bright in TeV energies (see Fig.9).
The electrons (positrons)
accelerated before at the reverse shock do not strongly influenced by synchrotron losses
in the weak magnetic field of the central part of the SNR and emit IC gamma-rays.

\section{Conclusion}

The main purpose of the present paper was the presentation of a
new numerical code for the modeling of hydrodynamics and nonlinear
shock acceleration in SNR. The code is described in details in
five Appendixes. It develops the modeling of particle acceleration
by spherical shocks fulfilled earlier by other authors
\cite{berezhko94,kang06}. Besides some important technical
details, the main novel features of our code include the particle
acceleration by two shocks - forward and reverse, the evolving
with a SNR age magnetic field, which determines the cosmic ray
transport, the account for the Alfv\'en drift effects both
upstream and downstream of the forward and reverse shocks. The
first version of the model of acceleration \cite{zirakashvili08b}
and a number of important applications including the explanation
of the overall spectrum of galactic cosmic rays \cite{ptuskin10},
and the modeling of particle acceleration and nonthermal radiation
in SNR RX J1713.7-3946 \cite{zirakashvili10} were developed upon
our code refinements.

The special attention in the astrophysical applications of the
 modeling discussed in the present work was focused on the particle
acceleration by the reverse SNR shock. It was shown that the reverse
shock can give a non-negligible contribution to the production of CR
ions and positrons as compared to the contribution of the forward shock.
The spectra of particles accelerated at the reverse shock can be harder
than the spectra at the forward shock, see Figures 8 and 10. It may offer a new
interpretation of cosmic ray data \cite{adriani09} that suggests the presence of
high energy primary positrons. The acceleration of positrons by the
reverse shock moving through the ejecta material, which contains the
positrons from $^{44}$Ti radioactive decays, was proposed in \cite{zirakashvili10b}. It is
also possible \cite{ptuskin11} that the hard spectrum of accelerated nuclei and
the low abundance of hydrogen in the material of supernova ejecta
accelerated by the reverse shock may explain the difference in the
energy spectra of hydrogen and helium in cosmic rays. We plan
detailed study of all these effects in a future work.

In the light of the problem of electron injection it is of interest
that the models of suprathermal electron injection \cite{zirakashvili10b,morlino11} reproduce
 the required amount of Galactic CR electrons and positrons if the leptons
are pre-accelerated up to $E_{inj} \sim 100$ MeV in the upstream regions of
supernova shocks.

The work was supported by the Russian Foundation for Basic Research grant 10-02-00110a.
We thank the anonymous referee for a number of valuable comments. 


\appendix
\onecolumn

\section{Details of the numerical method}

We used the following change of variables upstream of the shocks:

\begin{equation}
\xi =
\left\{ \begin{array}{ll}
\frac {r}{R_b}, \ r<R_b \\
\frac r{R_f}, \ r>R_{f}.
\end{array} \right. ,
\end{equation}

and downstream of the shocks:

\begin{equation}
\eta =
\left\{ \begin{array}{ll}
\frac {r-R_c}{R_c-R_b}, \ R_b<r<R_c, \\
\frac {r-R_c}{R_f-R_c}, \ R_c<r<R_{f}.
\end{array} \right.
\end{equation}

We shall use the dimensionless parameters $\tilde{t}=\ln (R_f/R_0)$
instead of time $t$ and $\zeta =\ln (p/mc)$ instead of momentum
$p$. It is convenient to use the new variable $n(\zeta )=4\pi
cp^4N(p)$ instead of momentum distribution $N(p)$. For the
relativistic momenta $p>>mc$ this variable corresponds to a
partial energy density of cosmic rays.

In these new variables the equations (1)-(4) have the following form upstream of the shocks:

\begin{equation}
\frac {\partial }{\partial \tilde{t}}R_{f,b}^3\xi ^2\rho =
-\frac {\partial }{\partial \xi }R_{f,b}^2\xi ^2(u-V_{f,b}\xi )\rho \frac {R_f}{V_f},
\end{equation}

\begin{equation}
\frac {\partial }{\partial \tilde{t}}R_{f,b}^3\xi ^2\rho u=
-\frac {\partial }{\partial \xi }R_{f,b}^2\xi ^2(u-V_{f,b}\xi )\rho u\frac {R_f}{V_f}-
R_{f,b}^2\xi ^2\frac {R_f}{V_f}\frac {\partial }{\partial \xi }(P_c+P_g),
\end{equation}


\begin{equation}
\frac {\partial }{\partial \tilde{t}}R_{f,b}^3\xi ^2P_g =
-\frac {\partial }{\partial \xi }R_{f,b}^2\xi ^2(u-V_{f,b}\xi )P_g \frac {R_f}{V_f}-
(\gamma _g-1)R_{f,b}^2\frac {R_f}{V_f}
\left[
(u-w)\xi ^2\frac {\partial }{\partial \xi }P_c+P_g\frac {\partial }{\partial \xi }\xi ^2u
\right],
\end{equation}

\begin{equation}
R_{f,b}^3\xi ^2\frac {V_f}{R_f} \frac {\partial n}{\partial \tilde{t}}=
\frac {\partial }{\partial \xi }
R_{f,b}\xi ^2D
\frac {\partial n}{\partial \xi }
-R_{f,b}^2\xi ^2(w-V_{f,b}\xi )\frac {\partial n}{\partial \xi }
+
\frac {R_{f,b}^2}3
\left[ \frac {\partial n}{\partial \zeta }-4n\right]
\frac {\partial }{\partial \xi }\xi ^2w,
\end{equation}

Here indexes $f$ and $b$ are referred to the forward ($\xi >1$)
and reverse shock ($0<\xi <1$) respectively.

The equations (1)-(4) have the following form downstream of the shocks:

\begin{equation}
\frac {\partial }{\partial \tilde{t}}r^2\Delta _{f,b}\rho =
-\frac {\partial }{\partial \eta }r^2u^{f,b}\rho \frac {R_f}{V_f}.
\end{equation}

\begin{equation}
\frac {\partial }{\partial \tilde{t}}r^2\Delta _{f,b}\rho u=
-\frac {\partial }{\partial \eta }r^2\frac {R_f}{V_f}(P_g+u^{f,b}\rho u)+
r^2\frac {R_f}{V_f}
\left(
\frac {2P_g\Delta _{f,b}}{r}-\frac {\partial }{\partial \eta }P_c
\right) ,
\end{equation}

\begin{equation}
\frac {\partial }{\partial \tilde{t}}r^2\Delta _{f,b}e =
-\frac {\partial }{\partial \eta }r^2\frac {R_f}{V_f}(uP_g+u^{f,b}e)
-wr^2\frac {R_f}{V_f}\frac {\partial }{\partial \eta }P_c.
\end{equation}

\begin{equation}
r^2\Delta _{f,b}\frac {V_f}{R_f}\frac {\partial n}{\partial \tilde{t}} =
\frac {\partial }{\partial \eta }
\frac {r^2D}{\Delta _{f,b}}\frac {\partial n}{\partial \eta }
-r^2w^{f,b}\frac {\partial n}{\partial \eta }
+
\frac {1}{3}
\left[ \frac {\partial n}{\partial \zeta } -4n\right] \frac {\partial }{\partial \eta }r^2w,
\end{equation}

Here the speeds $u^f$ and $u^b$ are determined as
$u^f=u-V_c(1-\eta )-\eta V_f$ and $u^b=u-V_c(1+\eta )+\eta V_b$ respectively.
The corresponding CR advective speeds $w^f$ and $w^b$ are determined as
$w^f=w-V_c(1-\eta )-\eta V_f$ and $w^b=w-V_c(1+\eta )+\eta V_b$ respectively.
The quantities $\Delta _{f,b}$ are the distances between the
forward shock and the contact discontinuity $\Delta _f=R_f-R_c$
and between the reverse shock and the the contact discontinuity
$\Delta _b=R_c-R_b$. We introduced energy density of the gas
$e=\frac {\rho u^2}{2}+\frac {P_g}{\gamma _g-1}$. Then the
equations (A7)-(A9) are written in the conservative form that is
convenient for the numerical method.  The radius $r$ in the last
equations should be expressed using Eq. (A2):

\begin{equation}
r =
\left\{ \begin{array}{ll}
R_c+\eta  (R_f-R_c), \ 0<\eta <1, \\
R_c+\eta  (R_c-R_b), \ -1<\eta <0.
\end{array} \right.
\end{equation}

\section{Solution of hydrodynamic equations in the upstream regions}

Since the flow upstream of the shocks is supersonic the hydrodynamic
equations can be solved using the following implicit numerical
scheme in these regions. Let introduce the grid $\xi _i$.
Integration of equations (A3)-(A5) on $\xi $ from $\xi _i$ to $\xi
_{i+1}$ results in the following expressions for density  $\rho
_i$, velocity $u_i$ and the gas pressure $P_{g,i}$ upstream of the
forward shock ($\xi >1$) at the grid knot with a number $i$  in
terms of $\rho _{i+1}$ and $u_{i+1}$:

\begin{equation}
u_i G_i=\rho _i^{old}u_i^{old}\left( \frac {R_f^{old}}{R_f}\right) ^3-
\rho _{i+1}u_{i+1}
\frac {3\tau \xi _{i+1}^2(u_{i+1}-V_f\xi _{i+1})}{V_f(\xi _{i+1}^3-\xi _i^3)}-
\frac {P_{c,i+1}+P_{g,i+1}^{old}-P_{c,i}-P_{g,i}^{old}}{V_f(\xi _{i+1}-\xi _i)}\tau
\end{equation}

\begin{equation}
\rho _i =G_i \left[ 1-\frac {3\tau \xi _{i}^2(u_i-V_f\xi _i)}{V_f(\xi _{i+1}^3-\xi _i^3)}\right] ^{-1}
\end{equation}

\[
P_{g,i}=
\left[
P_{g,i}^{old}\left( \frac {R_f^{old}}{R_f}\right) ^3-
P_{g,i+1}
\frac {3\tau \xi _{i+1}^2(u_{i+1}-V_f\xi _{i+1})}{V_f(\xi _{i+1}^3-\xi _i^3)}
-(\gamma _g-1)(w_i-u_i)\tau \frac {P_{c,i+1}-P_{c,i}}{V_f(\xi _{i+1}-\xi _i)}
\right] \times
\]
\begin{equation}
\left[ 1-3\tau
\frac
{\xi _{i}^2(u_i-V_f\xi _i)+(\gamma _g-1)(\xi _i^2u_i-\xi _{i+1}^2u_{i+1})}
{V_f(\xi _{i+1}^3-\xi _i^3)}
\right] ^{-1}
\end{equation}

Here the superscript '$old$' is referred to quantities at the
previous time step and $\tau $ is the
step in the dimensionless time $\tilde{t}$. The quantity $G_i$ is given by the expression

\begin{equation}
G_i=\rho _i^{old}\left( \frac {R_f^{old}}{R_f}\right) ^3-
\rho _{i+1}
\frac {3\tau \xi _{i+1}^2(u_{i+1}-V_f\xi _{i+1})}{V_f(\xi _{i+1}^3-\xi _i^3)}
\end{equation}

In spite of implicity Eqs (B1-B3) are readily solved recurrently
from $i=i_{max}$ down to $i=f$ corresponding to the position of the
forward shock. The value of the forward shock radius $R_f$ is related with the
 forward shock radius $R_f^{old}$ at the previous time step as $R_f=R_f^{old}\exp \tau $ in
 Eqs (B1-B4).

Performing the integration of equations (A3)-(A5) on $\xi $ from
$\xi _{i-1}$ to $\xi _i$ upstream of the reverse shock we obtain
the expressions for for density, velocity $\rho _i$, $u_i$ and and
the gas pressure $P_{g,i}$ upstream of the reverse shock ($\xi
<1$) in terms of $\rho _{i-1}$ and $u_{i-1}$:

\begin{equation}
u_i H_i'=u_i^{old}+
u_{i-1}
\frac {R_f\tau (u_{i}-V_b\xi _{i})}{R_bV_f(\xi _{i}-\xi _{i-1})}-
\frac {R_f(P_{c,i}+P_{g,i}^{old}-P_{c,i-1}-P_{g,i-1}^{old})}{R_bV_f(\xi _i-\xi _{i-1})}\tau
\end{equation}

\begin{equation}
\rho _i =H_i
\left[ 1+
\frac {3R_f\tau \xi _{i}^2(u^{old}_i-V_b\xi _i)}{R_bV_f(\xi _{i}^3-\xi _{i-1}^3)}
\right] ^{-1}
\end{equation}

\[
P_{g,i}=
\left[
P_{g,i}^{old}\left( \frac {R_b^{old}}{R_b}\right) ^3+
P_{g,i-1}
\frac {3R_f\tau \xi _{i-1}^2(u_{i-1}-V_b\xi _{i-1})}{R_bV_f(\xi _{i}^3-\xi _{i-1}^3)}
-(\gamma _g-1)(w_i-u_i)\tau \frac {R_f(P_{c,i}-P_{c,i-1})}{R_bV_f(\xi _{i}-\xi _{i-1})}
\right] \times
\]
\begin{equation}
\left[ 1+3R_f\tau
\frac
{\xi _{i}^2(u_i-V_b\xi _i)+(\gamma _g-1)(\xi _i^2u_i-\xi _{i-1}^2u_{i-1})}
{R_bV_f(\xi _{i}^3-\xi _{i-1}^3)}
\right] ^{-1}
\end{equation}

The quantities $H_i$ and $H'_i$ are given by expressions

\begin{equation}
H_i=\rho _i^{old}\left( \frac {R_b^{old}}{R_b}\right) ^3+
\rho _{i-1}
\frac {3R_f\tau \xi _{i-1}^2(u_{i-1}-V_b\xi _{i-1})}{R_bV_f(\xi _{i}^3-\xi _{i-1}^3)},
H'_i=1+
\frac {R_f\tau (u^{old}_i-V_b\xi _i)}{R_bV_f(\xi _{i}-\xi _{i-1})}
\end{equation}
Eqs (B5-B7) are solved recurrently from $i=1$ to $i=b$ corresponding to the position
of the reverse shock. The value of the new reverse shock radius $R_b$ in
 Eqs (B5-B8) was extrapolated using the
 reverse shock radius $R_b^{old}$ and velocity $V_b^{old}$ as $R_b=R_b^{old}+V_b^{old}R_f\tau /V_f$.

\section{Solution of hydrodynamic equations in the downstream regions}

Hydrodynamical quantities are determined at the centers of the cells with the indexes $i+1/2$ in the
 downstream regions.
We used an explicit numerical scheme of Trac \& Pen \cite{trac03}
for solution of Eqs. (A7)-(A9). In
accordance with these equations we use the following variables

\begin{equation}
U^1_{i+1/2}=\rho _{i+1/2}r^2_{i+1/2}\Delta _{f,b}, \
U^2_{i+1/2}=U^1_{i+1/2}u_{i+1/2}, \
U^3_{i+1/2}=r^2_{i+1/2}\Delta _{f,b}e_{i+1/2},
\end{equation}
and fluxes
\[
F^1_{i+1/2}=\frac {R_f}{V_f}
\rho _{i+1/2}r^2_{i+1/2}u^{f,b}_{i+1/2}, \
F^2_{i+1/2}=\frac {R_f}{V_f}
r^2_{i+1/2}(\rho _{i+1/2}u_{i+1/2}u^{f,b}_{i+1/2}+P_{g,i+1/2}),
\]
\begin{equation}
F^3_{i+1/2}=\frac {R_f}{V_f}
r^2_{i+1/2}(e_{i+1/2}u^{f,b}_{i+1/2}+u_{i+1/2}P_{g,i+1/2}).
\end{equation}

The variables $U^{\pm ,l}_{i+1/2}$ are introduced as
\begin{equation}
U^{\pm ,l}_{i+1/2}=c_sU^l_{i+1/2}\pm F^l_{i+1/2}, \ l=1,2,3
\end{equation}
Here $c_s$ is the so called freezing speed.

The values $U^{new, l}_{i+1/2}$ at
the next instant of time are given by

\begin{equation}
U^{new, 1}_{i+1/2}=U^{1}_{i+1/2}-\frac {\tau }{k_{f,b}}
(F^{1}_{i+1}-F^{1}_{i}),
\end{equation}
\begin{equation}
U^{new, 2}_{i+1/2}=U^{2}_{i+1/2}-\frac {\tau }{k_{f,b}}
(F^{2}_{i+1}-F^{2}_{i})
+\tau r_{i+1/2}\frac {R_f}{V_f}
\left(
2P_{g,i+1/2}\Delta _{f,b}-r_{i+1/2}\frac {P_{c,i+1}-P_{c,i}}{k_{f,b}}
\right) ,
\end{equation}
\begin{equation}
U^{new, 3}_{i+1/2}=U^{3}_{i+1/2}-\frac {\tau }{k_{f,b}}
(F^{3}_{i+1}-F^{3}_{i})
-\tau r^2_{i+1/2}w_{i+1/2}\frac {R_f}{V_f}
\frac {P_{c,i+1}-P_{c,i}}{k_{f,b}}
.
\end{equation}
Here $k_{f,b}$ is the grid step downstream of the forward and reverse shock respectively. Note
that  a uniform grid in the downstream region was used.

The fluxes $F^l_{i}$ at the grid knots between discontinuities are given by

\begin{equation}
F^l_{i}=0.5(F^{+,l}_{i}+F^{-,l}_{i}), \ l=1,2,3,
\end{equation}
where  the fluxes $F^{\pm ,l}_{i}$ at the
 grid knots are given by

\begin{equation}
F^{+,l}_{i}=U^{+,l}_{i-1/2}+0.5L(U^{+,l}_{i+1/2}-U^{+,l}_{i-1/2}, U^{+,l}_{i-1/2}-U^{+,l}_{i-3/2}),
\ l=1,2,3
\end{equation}

\begin{equation}
F^{-,l}_{i}=-U^{-,l}_{i+1/2}+0.5L(U^{-,l}_{i+1/2}-U^{-,l}_{i-1/2}, U^{-,l}_{i+3/2}-U^{-,l}_{i+1/2}),
 \ l=1,2,3.
\end{equation}
Here the function $L(a,b)$ is the nonlinear flux  limiter.

The numerical scheme is reduced to the first order Godunov's one for $L(a,b)=0$ in Eqs. (C8), (C9).
Such a scheme is very dissipative. The simplest second order scheme corresponds to $L(a,b)=a$.
However it is unstable. One should use more complex flux limiters for stability.
The different types of corresponding nonlinear limiters are available.
Trac \& Pen \cite{trac03} used a
so-called Van-Leer limiter:

\begin{equation}
L(a,b)=
\left\{ \begin{array}{ll}
\frac {2ab}{a+b}, \ ab>0 \\
0, \ ab\leqslant 0.
\end{array} \right. .
\end{equation}

We used a more dissipative "minmod" limiter:

\begin{equation}
L(a,b)=
\left\{ \begin{array}{ll}
a, \ ab>0, \ |a|<|b|, \\
b, \ ab>0, \ |a|>|b|, \\
0, \ ab\leqslant 0.
\end{array} \right. .
\end{equation}

For calculations of fluxes $F^{\pm ,l}_i$ near discontinuities we also use $L(a,b)=0$ or $L(a,b)=a$.

The fluxes $F^l_f$ and $F^l_b$  just downstream of the forward and reverse shock
at the knots with numbers $i=f$ and $i=b$
are calculated using hydrodynamical quantities
 just upstream of the forward and reverse shocks:

\[
F^1_{f,b}=\frac {R_f}{V_f}
\rho _{f,b}R^2_{f,b}(u_{f,b}-V_{f,b}), \
F^2_{f,b}=\frac {R_f}{V_f}
R^2_{f,b}(\rho _{f,b}u_{f,b}(u_{f,b}-V_{f,b})+P_{g,f,b}),
\]
\begin{equation}
F^3_{f,b}=\frac {R_f}{V_f}
R^2_{f,b}(e_{f,b}(u_{f,b}-V_{f,b})+u_{f,b}P_{g,f,b}).
\end{equation}

The fluxes $F^l_c$  at the position of the contact discontinuity
 at the knot with number $i=c$ are given by

\[
F^1_{c}=0, \ F^2_c=P'R^2_cR_f/V_f, \ F^3_c=V_cF^2_c,
\]
where the pressure $P'$  is found from the approximate solution of the Riemann problem for decay of an
 arbitrary discontinuity:

\begin{equation}
P'=\frac
{P_{c+1/2}\sqrt{\rho_{c-1/2}}+P_{c-1/2}\sqrt{\rho_{c+1/2}}+(u_{c-1/2}-u_{c+1/2})
\sqrt{\gamma _gP\rho _{c-1/2}\rho _{c+1/2}}}
{\sqrt{\rho_{c-1/2}}+\sqrt{\rho_{c+1/2}}}.
\end{equation}
Here $P=(P_{c-1/2}+P_{c+1/2})/2$.

Hydrodynamical quantities at the new time step
 are found using Eq. (C1) after the calculation of new positions of the
 discontinuities (see Appendix D).

For a stability of the numerical scheme considered the freezing speed $c_s$ must be greater than the
maximal sonic velocity. We use the following expression for $c_s$

\begin{equation}
c_s=\frac {R_f}{V_f}
\max{\left[
\left(
\sqrt{\gamma _g\frac {0.5(P_{c,i}+P_{c,i+1})+P_{g,i+1/2}}{\rho _{i+1/2}}}+|u_{i+1/2}|
\right) \Delta _{f,b}^{-1}
\right] }, R_b<r_{i+1/2}<R_f
\end{equation}

The time step $\tau $ was found from the relation $\tau =0.5\max (k_f,k_b)/c_s $.

\section{Calculation of the velocities of discontinuities}

The velocities of discontinuities are found from the approximate solutions of the Riemann problem describing
a decay of an arbitrary discontinuity.
For the forward and reverse shocks the solution for the shock velocities $V_{f}$ and $V_{b}$ are the following:

\begin{equation}
V_{f}=u_{f}+ \rho ^{-1/2}_{f}\sqrt{\frac {\gamma _g+1}{2}P_{f-1/2}+\frac {\gamma _g-1}{2}P_{f}}.
\end{equation}

\begin{equation}
V_{b}=u_{b}- \rho ^{-1/2}_{b}\sqrt{\frac {\gamma _g+1}{2}P_{b+1/2}+\frac {\gamma _g-1}{2}P_{b}}.
\end{equation}

The shock velocities are determined by upstream values $P_f, P_b, u_f, u_b, \rho _u, \rho _b$
and downstream  values
$P_{f-1/2}, P_{b+1/2}$ of the gas pressure, velocity and density.

The expression for the velocity of the contact discontinuity $V_c$ is given by

\begin{equation}
V_c=\frac {u_{c-1/2}\sqrt{\rho_{c-1/2}}+u_{c+1/2}\sqrt{\rho_{c+1/2}}
+\frac {P_{c-1/2}-P_{c+1/2}}{\sqrt{\gamma _gP}}}{\sqrt{\rho_{c-1/2}}+\sqrt{\rho_{c+1/2}}}
\end{equation}
Here $P=(P_{c-1/2}+P_{c+1/2})/2$.

The new values for the radii $R_b$, $R_c$, $R_f=R^{old}_f\exp (\tau )$ and real time $t$
are now found:

\begin{equation}
R_b=R_b^{old}+R_f^{old}\tau \frac {V_b+V_b^{old}}{V_f+V_f^{old}}\exp (0.5\tau ),
\end{equation}
\begin{equation}
R_c=R_c^{old}+R_f^{old}\tau \frac {V_c+V_c^{old}}{V_f+V_f^{old}}\exp (0.5\tau ),
\end{equation}
\begin{equation}
t=t^{old}+\frac {2R_f^{old}\tau }{V_f+V_f^{old}}\exp (0.5\tau ).
\end{equation}

\section{Solution of CR transport equation}

The finite difference scheme for Eqs (A6) and (A10) can be written
in the following form:

\begin{equation}
a_i^jn_{i-1}^j+b_i^jn_i^j+c_i^jn_{i+1}^j=g_i^j
\end{equation}

Here $n_i^j$ are the values of $n$ at the new time step calculated in the grid knot $\xi _i$ or $\eta _i$ and
at $\zeta _j$. We use the following coefficients upstream of the  shocks

\begin{equation}
a_i^j=D_{i-1/2}^jR_{f,b}\frac {\xi ^2_{i-1/2}}{\xi _{i}-\xi _{i-1}}+
(w_i-V_{f,b}\xi _i)R_{f,b}^2\xi _i^2
\left\{ \begin{array}{ll}
0, \ \xi _i> 1, \\
1, \ \xi _i<1.
\end{array} \right. ,
\end{equation}
\begin{equation}
c_i^j=D_{i+1/2}^jR_{f,b}\frac {\xi ^2_{i+1/2}}{\xi _{i+1}-\xi _{i}}-(w_i-V_{f,b}\xi _i)R_{f,b}^2\xi _i^2
\left\{ \begin{array}{ll}
1, \ \xi _i>1, \\
0, \ \xi _i< 1.
\end{array} \right. ,
\end{equation}

\begin{equation}
b_i^j=-a_i^j-c_i^j-\frac {V_fR^3_{f,b}}{3R_f\tau }(\xi ^3_{i+1/2}-\xi ^3_{i-1/2})-\frac {d_iR^2_{f,b}}3
\left\{ \begin{array}{ll}
4-h^{-1}, \ d_i\leqslant 0, \\
4, \ d_i>0.
\end{array} \right. , \
d_i=\xi _{i+1/2}^2w_{i+1/2}-\xi _{i-1/2}^2w_{i-1/2},
\end{equation}

\begin{equation}
g_i^j=-\frac {V_fR^3_{f,b}}{3R_f\tau }(\xi ^3_{i+1/2}-\xi ^3_{i-1/2})n^{j,old}_i-\frac {d_iR^2_{f,b}}{3h}
\left\{ \begin{array}{ll}
-n^{j-1}_i, \ d_i\leqslant 0, \\
n^{j+1,old}_i-n^{j,old}_i, \ d_i>0.
\end{array} \right. \ .
\end{equation}

The corresponding coefficients downstream of the shocks are given by
\begin{equation}
a_i^j=D_{i-1/2}^j\frac {r_{i-1/2}^2}{\Delta _{f,b}(\eta _{i}-\eta _{i-1})}+
w^{f,b}_{i}r_i^2
\left\{ \begin{array}{ll}
0, \ \eta _i>0, \\
1, \ \eta _i<0.
\end{array} \right. ,
\end{equation}
\begin{equation}
c_i^j=D_{i+1/2}^j\frac {r_{i+1/2}^2}{\Delta _{f,b}(\eta _{i+1}-\eta _{i})}-w^{f,b}_{i}r_i^2
\left\{ \begin{array}{ll}
1, \ \eta _i>0, \\
0, \ \eta _i<0.
\end{array} \right. ,
\end{equation}

\begin{equation}
b_i^j=-a_i^j-c_i^j-\frac {V_f}{3R_f\tau }(r^3_{i+1/2}-r^3_{i-1/2})-\frac {d_i}3
\left\{ \begin{array}{ll}
4-h^{-1}, \ d_i\leqslant 0, \\
4, \ d_i>0.
\end{array} \right. , \
d_i=r_{i+1/2}^2w_{i+1/2}-r_{i-1/2}^2w_{i-1/2},
\end{equation}

\begin{equation}
g_i^j=-\frac {V_f}{3R_f\tau }(r^3_{i+1/2}-r^3_{i-1/2})n^{j,old}_i-\frac {d_i}{3h}
\left\{ \begin{array}{ll}
-n^{j-1}_i, \ d_i\leqslant 0, \\
n^{j+1,old}_i-n^{j,old}_i, \ d_i>0.
\end{array} \right. \ .
\end{equation}

The forward and reverse shocks are situated at knots with numbers $i=f$ and $i=b$ respectively.
The coefficients at these knots are given by

\begin{equation}
a_f^j=D_{f-1/2}^j\frac {r_{f-1/2}^2}{\Delta _{f}(\eta _{f}-\eta _{f-1})}, \
c_f^j=D_{f+1/2}^jR_{f}\frac {\xi ^2_{f+1/2}}{\xi _{f+1}-\xi _{f}},
\end{equation}

\begin{equation}
b_f^j=-a_f^j-c_f^j-\frac {V_f}{3R_f\tau }(R^3_f\xi ^3_{f+1/2}-r^3_{f-1/2})-\frac {4-h^{-1}}3d_f, \
d_f=R_f^2\xi _{f+1/2}^2w_{f+1/2}-r_{f-1/2}^2w_{f-1/2},
\end{equation}

\begin{equation}
g_f^j=-\frac {V_f}{3R_f\tau }(R_f^3\xi ^3_{f+1/2}-r^3_{f-1/2})n^{j,old}_f
-\eta ^f(V_f-w_f)\rho _f\frac {p^f_{inj}c}{mh}\delta (j,j_f)
+\frac {d_f}{3h}n^{j-1}_f,
\end{equation}

and

\begin{equation}
a_b^j=D_{b-1/2}^jR_{b}\frac {\xi ^2_{b-1/2}}{\xi _{b}-\xi _{b-1}},
c_b^j=D_{b+1/2}^j\frac {r_{b+1/2}^2}{\Delta _{b}(\eta _{b+1}-\eta _{b})},
\end{equation}

\begin{equation}
b_b^j=-a_b^j-c_b^j-\frac {V_f}{3R_f\tau }(r^3_{b+1/2}-R^3_b\xi ^3_{b-1/2})-\frac {4-h^{-1}}3d_b
, \
d_b=r_{b+1/2}^2w_{b+1/2}-R_b^2\xi _{b-1/2}^2w_{b-1/2},
\end{equation}

\begin{equation}
g_b^j=-\frac {V_f}{3R_f\tau }(r^3_{b+1/2}-R_b^3\xi ^3_{b-1/2})n^{j,old}_b
-\eta ^b(w_b-V_b)\rho _b\frac {p^b_{inj}c}{mh}\delta (j,j_b)
+\frac {d_b}{3h}n^{j-1}_b \ .
\end{equation}
Here $\delta (j,l)$ is the Kronecker's symbol while $j_f$ and $j_b$
are numbers of the momentum grid knots corresponding to injection
momenta $p_f$ and $p_b$ respectively.

The contact discontinuity is situated at the knot $i=c$. The
coefficients at this knot are given by

\begin{equation}
a_c^j=D_{c-1/2}^j\frac {r_{c-1/2}^2}{\Delta _{b}(\eta _{c}-\eta _{c-1})}, \
c_c^j=D_{c+1/2}^j\frac {r_{c+1/2}^2}{\Delta _{f}(\eta _{c+1}-\eta _{c})},
\end{equation}

\begin{equation}
b_c^j=-a_c^j-c_c^j-\frac {V_f}{3R_f\tau }(r^3_{c+1/2}-r^3_{c-1/2})-\frac {d_c}3
\left\{ \begin{array}{ll}
4-h^{-1}, \ d_c\leqslant 0 \\
4, \ d_c>0
\end{array} \right. , \
d_c=r_{c+1/2}^2w_{c+1/2}-r_{c-1/2}^2w_{c-1/2},
\end{equation}

\begin{equation}
g_c^j=-\frac {V_f}{3R_f\tau }(r^3_{c+1/2}-r^3_{i-1/2})n^{j,old}_c-\frac {d_c}{3h}
\left\{ \begin{array}{ll}
-n^{j-1}_c, \ d_c\leqslant 0 \\
n^{j+1,old}_c-n^{j,old}_c, \ d_c>0
\end{array} \right. \ .
\end{equation}

We use a standard method to solve the tridiagonal set of equations
(E1) (e.g. Godunov \cite{godunov71}). At first the following
coefficients $L^j_{i+1/2}$ and  $K^j_{i+1/2}$ are recurrently
calculated from $i=1$ up to $i=i_{max}-1$:

\begin{equation}
L^j_{i+1/2}=-\frac {c^j_i}{a^j_iL^j_{i-1/2}+b^j_i},
\ K^j_{i+1/2}=\frac {g^j_{i}-a^j_iK^j_{i-1/2}}{a^j_iL^j_{i-1/2}+b^j_i}.
\end{equation}
The values $L^j_{1/2}=1$ and $K^j_{1/2}=0$ correspond to the boundary
condition $n^j_0=n^j_1$ at the center of the remnant. The
functions $n^j_i$ are calculated recurrently from $i=i_{max}-1$
down to $i=0$ as

\begin{equation}
n^j_i=L^j_{i+1/2}n^j_{i+1}+K^j_{i+1/2}.
\end{equation}
We put $n^j_i=0$ at $i=i_{max}$. This describes an absorbtion of
particles at the boundary of the simulation domain. The equations
(E19), (E20) are recurrently used for all values of $j$ starting
with the lowest energies at $j=j_{min}+1$.

\section{Initial conditions and the code performance}

We use a uniform spacial grid downstream of the shocks and the
following non-uniform grids upstream of the forward shock

\begin{equation}
\xi _i=1-\xi _{min}+\xi _{min}\exp [(i-f)k_1],\ k_1=\ln [(\xi
_{max}+\xi _{min}-1)/\xi _{min}]/(i_{max}-f)
\end{equation}

and upstream of the reverse shock

\begin{equation}
\xi _i=2ik_4+\left[ \ln \left( 1+(2\xi _{min})^{(1-2ik_4)}\right) -\ln (1+2\xi _{min})\right] /\ln (2\xi _{min}),
\ k_4=1/b.
\end{equation}
Here the parameter $\xi _{max}$ describes the ratio of radii of the simulation domain and forward shock.
This parameter
is constant during the simulation. We use the value $\xi _{max}=2$.
The value of the parameter $\xi _{min}\sim 10^{-11}$ should be small enough to resolve
 the spacial variation scale of the lowest energy cosmic rays upstream of the shocks.

We use the following profiles of the gas density, velocity and
pressure at the initial moment of time $\tilde{t}=0$:

\begin{equation}
\left\{ \begin{array}{ll}
\rho =\rho _{0},\ u=0,\ P_g=P_{0},\ r>R_f, \\
\rho =4\rho _{0},\ u=0.75V_f,\ P_g=0.75\rho _0V_f^2, \ R_b<r<R_f, \\
\rho =\rho _{0}(r/R_b)^{-k},\ u=1.5V_fr/R_b,\ P_g=0,\ r_{ej}<r<R_b, \\
\rho =\rho _{0}(r_{ej}/R_b)^{-k},\ u=1.5V_fr/R_b, P_g=0,\
r<r_{ej}.
\end{array} \right.
\end{equation}
Here $\rho_0$ and $P_0$ are the gas density and pressure in the
circumstellar medium while $r_{ej}=2V_{ej}R_b/3V_f$ is the initial
radius of the flat part of the ejecta density distribution. The
initial velocities of the reverse shock and contact discontinuity
are $V_b=0.5V_f$ and $V_c=0.75V_f$ respectively. Their initial
radii can be chosen rather arbitrary. We use values $R_b=0.9R_f$
and $R_c=0.95R_f$.

The initial cosmic ray number density is zero. At every time step the system of equations (1)-(4) is solved in
the following order:

1) The hydrodynamical equations are solved in the upstream regions
(see Appendix B).

2) These equations are solved in the downstream regions
and the speeds of discontinuities are determined (see
Appendix C and D).

3) The parameters of CR transport depending on the hydrodynamical
quantities are calculated.

4) CR transport equation is solved (see appendix E). Cosmic ray
pressure is calculated.

It is important to note that we use the finite difference method
for solution of CR transport equations even at the shock positions
at grid knots $i=f$ and $i=b$ (see Eqs. (E10-E15)). These
equations contain terms originating from the first derivative on
time (see terms containing $\tau $ in Eqs.(E10-E15)). These terms
will not appear if we use exact boundary conditions at the shock fronts
which can be derived by integration of Eq. (4) in the vicinity
of the shocks. Our method permits to use relatively large time
steps $\tau $ while
 the use of the exact boundary conditions results in a numerical instability for the large time steps.
Physically this is because the low energy particles have very
small acceleration times in the case of Bohm-like diffusion. This
results in the numerical instabilities. It is possible to use many
iterations
 at every time step \cite{berezhko94} or to use very small time steps
near the shock front \cite{kang06} to avoid this
problem. This results in significantly longer times of
simulations.

The numerical scheme with Eqs. (E10-E15) artificially increases
acceleration time of low-energy particles and makes the
calculations to be stable. This means that our method can not be
used for simulations of fast varying processes. However if the
shocks propagate in the smooth environment then the low-energy
particles are accelerated in a quasi-steady regime and our method
is well justified.

The numerical results shown in Figs (1)-(10) were obtained
 using the grid with 200$\times $200 radius-momentum cells  in the every of four regions
 (two upstream and
two downstream regions of the shocks). The initial radius of the flat part of the ejecta density
 distribution $r_{ej}$ is $r_{ej}=R_b/6$.
The calculation
 of the remnant evolution takes two hours at PC. The total energy
 is conserved with 5\% accuracy. Preliminary results can be obtained using ten times faster
 crude calculations with 100$\times $100 grid.












\end{document}